\documentclass[12pt,a4paper]{article}
\pdfoutput=1
\usepackage{cite}
\usepackage{jheppub}
\usepackage{amsmath,amsthm,amssymb,graphicx,slashed}
\usepackage{booktabs}
\usepackage{hyperref}
\usepackage[usenames,dvipsnames]{xcolor}

\numberwithin{equation}{section}

\def\p{\partial}

\def\be{\begin{equation}}
\def\ee{\end{equation}}
\def\ba{\begin{align}}
\def\ea{\end{align}}
\def\beq{\begin{eqnarray}}
\def\eeq{\end{eqnarray}}
\def\a{\alpha}

\newcommand\eref[1]{\eqref{#1}}

\title{The Hot Attractor Mechanism:\\ Decoupling Without Deep Throats}

\author[*]{Kevin Goldstein,}
\author[*]{Vishnu Jejjala,}
\author[*,\dagger]{Suresh Nampuri}

\affiliation[*]{Mandelstam Institute for Theoretical Physics, School of Physics, and National Institute for Theoretical Physics, University of the Witwatersrand, Johannesburg, WITS 2050, South Africa}
\affiliation[\dagger]{Center for Mathematical Analysis, Geometry and Dynamical Systems, Instituto Superior T\'ecnico, Universidade de Lisboa, Av.\ Rovisco Pais, 1049-001 Lisboa, Portugal}

\emailAdd{kevin.goldstein@wits.ac.za}
\emailAdd{vishnu@neo.phys.wits.ac.za}
\emailAdd{snampuri@math.tecnico.ulisboa.pt}

\abstract{
Non-extremal black holes in $\mathcal{N}=2$ supergravity have two horizons, the geometric mean of whose areas recovers the horizon area of the extremal black hole obtained from taking a smooth zero temperature limit.
In prior work~\cite{1410.3478} using the attractor mechanism, we deduced the existence of several  moduli independent invariant quantities obtained from averaging over a decoupled inter-horizon region.
We establish that non-extremal geometries at the Reissner--Nordstr\"om point, where the scalar moduli are held fixed, can be lifted to solutions in supergravity with a near-horizon AdS$_3\times$S$^2$.
These solutions have the same entropy and temperature as the original black hole and therefore allow an interpretation of the underlying gravitational degrees of freedom in terms of CFT$_2$.
Symmetries of the moduli space enable us to explicate the origin of entropy in the extremal limit.
}

\begin{document}

\maketitle

\section{Introduction}
Decoupling is crucial to the holographic correspondence.
The equivalence between string theory on AdS$_5$ and four dimensional ${\cal N}=4$ super-Yang--Mills conformal field theory (CFT) constitutes the paradigmatic example~\cite{adscft1,adscft2,adscft3}.
The gauge theory lives on the worldvolume of a stack of D$3$-branes, or equivalently at the boundary of AdS$_5$.
String modes in the near-horizon geometry get stuck in the throat and are gravitationally bound to the branes whose backreaction sources the AdS$_5\times$S$^5$ spacetime.
The ten dimensional flat space supergravity decouples, and we can investigate the remainder of the theory encapsulating the D-brane dynamics either in terms of an open string language or a dual closed string language.
This decoupling essentially isolates the gravitational degrees of freedom of relevance.

The attractor mechanism for extremal black holes in ${\cal N}=2$ supergravity constitutes another example of the same phenomenon~\cite{attr1,attr2,attr3,attr4,hep-th/0506177,hep-th/0606244,hep-th/0611143,0708.1270}.
The entropy of a black hole enumerates the number of microstates, which is an integer that is determined by the quantized charges.
The moduli at asymptotic infinity, however, vary continuously.
The horizon area and thus the entropy must be independent of the background moduli, which then assume fixed values at the horizon.
The reason this happens is that the radial direction corresponds to an infinitely long AdS$_2$ throat.
Fluctuations in the asymptotic values of the scalar fields damp down inside the throat, whereas the horizon values, which are fixed by the charges, are unchanged.

As we expect, because of holography, the number of degrees of freedom of the system is codimension one in gravity~\cite{ts1,ts2}.
While in the most symmetric examples, we can promote gravitational thermodynamics to statistical physics in the microcanonical ensemble~\cite{math1,math2,skta,benawarner,llm,bdjs,bdms,js} and in fact map the precise evolution of states from weak coupling to strong coupling~\cite{sv1,sv2,sv3,sv4}, we do not have the ability to identify the microstates that account for entropy in generic examples.
The most interesting cases about which least is known are the non-extremal black holes, which have a finite temperature (surface gravity) associated to the event horizon.
The explicit construction of regular solutions in supergravity with the same charges as non-extremal black holes provides validation for this program~\cite{me,bgrw,ccdm,dgr,bnw,nie,ctv,Bena:2015drs} and has led to the determination of new decoupling limits~\cite{bdjs2,fgms}.

A curious property of non-extremal black holes is that the product of the area of the horizons is a power of the area of the extremal black hole obtained from taking the zero temperature limit~\cite{finn,cgp,cretc1}.
In particular, for the Reissner--Nordstr\"om geometry, the extremal black hole's horizon area is the geometric mean of the areas of the event horizon and the inner (Cauchy) horizon for the various representatives of a family of non-extremal black holes each of which is at a different temperature:
\be
A_\mathrm{ext} = \sqrt{A_+ A_-} ~. \label{eq:gm}
\ee
In our previous paper on this subject~\cite{1410.3478}, we proved this relation for a large class of black holes in ${\cal N}=2$ supergravity by exploiting the attractor mechanism.

In this article, we take initial steps toward understanding the organization of microstates for non-extremal black holes in the context of the four dimensional ungauged ${\cal N}=2$ supergravity that arises in the low energy limit of type II string compactifications.
The action determining the dynamics of field configurations in this theory is Einstein gravity coupled to neutral scalars and $U(1)$ gauged fields.
Black hole backgrounds in this theory, at fixed charges, are characterized by the asymptotic values of the scalar fields whose flow is governed by a potential that is a fixed function of the charges and scalars.
The Lorentz-violating vacuum solutions in this theory that extremize the classical action are zero temperature black holes, whose horizon acts as an attractor for the scalar flows.
At the attractor point, the scalars reside at the minimum of the potential.\footnote{
Modulo flat directions, the potential has a unique minimum in the range of asymptotic scalar field values corresponding to the single centered black holes;
this minimum is a function of the charges of the black hole.}
The attractor near-horizon geometry is characterized by a single length scale corresponding to the extremum of the potential.
Due to the attractor mechanism, the values of scalar fields in the near-horizon geometry are independent of their asymptotic values.
The horizon area of the black hole sets the only length scale in the geometry and in consequence is related to the extremum of the potential and independent of the asymptotic scalar field values.
The attractor mechanism classifies the solution space of scalar flows in terms of fixed points and identifies the geometry whose asymptotic symmetry group governs the organization of microstates in the Hilbert space.
The dimension of the Hilbert space is then expressed as an invariant length scale of the geometry.
In considering a non-extremal black hole, we introduce a second length scale, the temperature, which we can express in terms of the proper distance between the two horizons, or the deviation from extremality.
It is the existence of this second length scale which explains~\eref{eq:gm}.

The organization of the paper is as follows.
In Section~\ref{sec:two}, we describe the setup and recapitulate the derivation of~\eref{eq:gm}.
At both the inner horizon and the outer horizon, the scalar fields assume values that depend on the asymptotic moduli, which serve as the initial condition for the flow.
Nevertheless, as we determined in~\cite{1410.3478}, there are robust invariants constructed from integrating over the region between the two horizons.
This is a decoupled region, and it is this decoupling that enables us to explore universal features of non-extremal solutions.
In Section~\ref{sec:three}, as a first step to providing a microscopic interpretation of degrees of freedom associated to the inter-horizon region, we construct a black hole with the same entropy and temperature as the original non-extremal black hole with the added feature that there is a manifest AdS$_3\times$S$^2$ description.
This facilitates an interpretation of the entropy in terms of CFT$_2$.
We as well investigate fluctuations in the moduli space of hot attractor geometries about the Reissner--Nordstr\"om solution and discover the symmetries of AdS$_2$.
In the extremal limit, we find that the AdS$_2$ symmetry in moduli space becomes a symmetry in spacetime.
Coupled with the hot attractor mechanism, which averages over the inter-horizon region, this enables us to identify degrees of freedom that contribute to the entropy.
In Section~\ref{sec:four}, we discuss our results, the limitations of the methods we employ, and compare to recent alternative approaches to a statistical description of the entropy of non-extremal black holes.
Finally, we offer a prospectus for future work.

\section{Setup and background}\label{sec:two}
\subsection{Equations of motion}
To establish notation and introduce important features of attractor spacetimes, we review the equations of motion for the action of interest. 
We consider four dimensional gravity coupled to $U(1)$ gauge fields and moduli,
\begin{equation}
  S=\frac{1}{\kappa^{2}}\int d^{4}x\sqrt{-G}(R-2 g_{ij}(\phi)\partial_\mu\phi^i \partial^\mu \phi^j-
  f_{ab}(\phi)F^a_{\mu \nu} F^{b \ \mu \nu} -{\textstyle{1 \over 2}} {\tilde f}_{ab}(\phi) F^a_{\mu \nu}
  F^b_{\rho \sigma} \epsilon^{\mu \nu \rho \sigma} ) ~.
  \label{actiongen}
\end{equation}
Taking a static spherically symmetric background geometry 
\begin{equation}
ds^2 = - a(r)^2\, dt^2 + a(r)^{-2}\, dr^2 + b(r)^2\, d\Omega_2^2\,,\label{MetAnsatz}
\end{equation}
and an appropriate ansatz for the gauge fields (see~\cite{hep-th/0507096}  for details),
we obtain the following equations of motion:\footnote{The effective potential, $V_\text{eff}$, is given by $V_\text{eff}(\phi)=f^{ab}(Q_{ea}-{\tilde f}_{ac}Q^c_m)(Q_{eb}- {\tilde f}_{bd}Q^d_m)+f_{ab}Q^a_mQ^b_m$.
The constants $Q_m^a$ and $Q_{ea}$ encode charges carried by the gauge fields and $f^{ab}$ is  the
matrix inverse of $f_{ab}$.}
\begin{eqnarray}
	(a^2 b^2)''&=&2~,\label{att1}\\
	\dfrac{b''}{b} &=&- (\phi')^2~,\label{att2}\\
		\left(a^2 b^2 (\phi_i)'\right)'&=& \dfrac{\partial _{\phi^i}V_\text{eff}(\phi)}{2 b^2}\label{att3}~,
\end{eqnarray}
together with  the Hamiltonian constraint 
\begin{equation}
	-1 + a^2 b'^2 + \frac12 (a^2)' (b^2)' = - \frac{V_\text{eff}}{b^2} + a^2 b^2 \phi'^2 ~, \label{constraint}
\end{equation}
where $\phi'^2$ is short hand for $g_{ij}{\phi^i}'{\phi^j}'$.
Using the equations of motion,~\eqref{constraint} can also be written in the following convenient form:
\begin{equation}
	\label{constraint2}
	1- \frac{(a^2 (b^2)')'}{2} = \frac{V_\text{eff}(\phi)}{b^2}~.
\end{equation}
We also note that (\ref{att1})--(\ref{att3}) follow from extremizing the effective $(1+1)$ dimensional action,
\begin{equation}
\label{eq:1D}
S= \int dt \int dr \left((a^2 b)'b' - a^2 b^2 g_{ij}{\phi'}^i{\phi'}^j - \frac{V_\text{eff}(\phi)}{b^2}  -1\right)~.
\end{equation}
with~\eqref{constraint} imposed as a constraint.

\subsection{Moduli spaces and attractors: Hot and cold}
In this section, we discuss attractive properties of both hot and cold black holes.
The attractor mechanism is a well known feature of extremal (\textit{i.e.}, zero temperature) black holes.
The scalar moduli flow to fixed values at the horizon so that the black hole entropy is independent of their asymptotic values.\footnote{
Not all scalars are necessarily fixed.
Flat directions --- \textit{i.e.}, scalars whose value on the horizon does not effect the entropy --- are not constrained.}
Originally viewed as a consequence of supersymmetry~\cite{attr1}, the attractor mechanism can be seen as simply a feature of double horizons which appear in the zero temperature limit as we review below~\cite{attr5,hep-th/0507096}.
Thus, it is \textit{extremality} that is the crucial point.
Starting with~\eqref{att1}, which is easily integrated, one obtains
\begin{equation}
\label{att4}
a^2 b^2=(r-r_+)(r-r_-)~.
\end{equation} 
Assuming that $b^2$ remains finite, the zeros of~\eqref{att4} correspond to two horizons.\footnote{
The inner horizon is thermodynamically unstable~\cite{penrose68,poisson90}.
Moreover, it has a negative temperature corresponding to the fact that as a black hole becomes more non-extremal --- \textit{i.e.}, the temperature of the outer horizon increases --- the inner horizon becomes smaller.
In the framework of supergravity, we regard the inner horizon as defining a thermodynamically interesting arena.}
Demanding periodicity of the Euclidean time direction at the outer horizon fixes 
\begin{equation}
T=\frac{(a^2)'_+ }{4\pi }=  \frac{(r_+ - r_-)}{4\pi b^2_+} = \frac{\Delta}{2\pi b^2_+}  ~, \label{eq:T}
\end{equation}
where $\Delta=\frac12(r_+-r_-)$, related to the coordinate distance between the horizons and often called the non-extremality parameter, characterizes the deviation from extremality.
From~\eqref{att4} and~\eqref{eq:T}, we see that in the limit $r_+\rightarrow r_-$, we have an extremal black hole with $a^2$ having a double zero at the horizon.
Now, evaluating~\eqref{att2} and~\eqref{constraint} at extremality, using the fact that $a^2$ has a double zero, gives the equations which encode the attractor mechanism: 
\begin{eqnarray}
\left.\frac{\partial_{\phi^i} V_\text{eff}(\phi)}{b^2}\right|_{r=r_\text{ext}} &=& 0\,,\label{A1}\\
 \left[\frac{V_\text{eff}(\phi)}{b^2}-1\right]_{r=r_\text{ext}}&=&0\label{A2}\,,
\end{eqnarray}
where $r_\text{ext}$ is the position of the double horizon.
Henceforth, we will denote any quantity evaluated  on the horizon of the extremal solution with the subscript ``ext.''
From~\eqref{A1} we see that the scalars are fixed at the minimum of the effective potential $V_\text{eff}$. The position of this minimum depends on the charges carried by the black hole but not on the asymptotic moduli.\footnote{
Though~\eqref{A1} only ensures that we are at an extremum of the potential, for a black hole solution to exist we need it to be at a minimum.}
Using the usual area entropy relation,~\eqref{A2} can be written 
\begin{equation}
S_\text{ext}=\frac{1}{4}A_\text{ext}= \pi b_\text{ext}^2  =\pi V_\text{ext}~,\label{Sext}
\end{equation}
so that entropy just depends on the minimum of the potential and not  the asymptotic moduli.  

As we have sen, for extremal black holes in asymptotically flat backgrounds there is only one length scale in the problem related to the horizon area.
This scale is fixed by (\ref{A1}) and (\ref{A2}).
For non-extremal black holes the temperature introduces a second length scale, $\Delta$, which, as one can see from~\eqref{eq:T}, can be related to the coordinate distance between the two horizons.
This suggests that any generalization of the attractor mechanism for non-extremal black holes should involve both horizons.
In fact in, as shown in~\cite{1410.3478}, by integrating~\eqref{att3} and~\eqref{constraint2} between the horizons, we generalize~\eqref{A1} and~\eqref{A2} to the averaged equations  
\begin{eqnarray}
\int^{r_+}_{r_- }dr\left(\frac{\partial_{\phi^i} V_\text{eff}(\phi)}{b^2}\right) &=& 0\,,\label{hot1}\\
\int^{r_+} _{r_-}dr \left(\frac{V_\text{eff}(\phi)}{b^2}-1\right)&=&0\,.\label{hot2}
\end{eqnarray}
In~\cite{1410.3478}, we show that~\eqref{Sext}, which tells us that the area of the extremal black hole is independent of the asymptotic moduli also generalizes.
Indeed, the product of the horizon areas is the quantity which is independent of the asymptotic moduli and therefore related to the area of the extremal solution:
\begin{equation}
A_+ A_- = A_\text{ext}^2~,\label{AreaLaw}
\end{equation}
where $A_\pm$ are the areas of the inner and outer horizons respectively and $A_\text{ext}$ is the extremal area.
Using the usual proportionality relation between area and entropy~\eqref{AreaLaw} becomes 
\begin{equation}
S_+ S_- = S_\text{ext}^2~,\label{EntropyLaw}
\end{equation}
though it is not, a priori, clear how $S_-$ should be interpreted.
What is clear is that $S_\pm$ both depend on the asymptotic values of the moduli but $S_\text{ext}$ does not.
This means that there is some non-trivial cancellation involving both horizons. 

We now present an alternative derivation of the area law to the one discussed in~\cite{1410.3478}.
Non-extremal solutions, can be regarded as thermal excitations above the extremal backgrounds.
While  extremal solutions (in asymptotically flat spacetime) are characterized by only one length scale, namely the circumferential radius of the horizon, $b_\text{ext}$, turning on a temperature introduces a new length scale --- the non-extremality parameter, $\Delta$.
There are two other length scales which are natural in the non-extremal solution, \textit{viz.}, the  circumferential radii of the inner and outer horizons, $b_-$ and $b_+$, respectively. 
Given that, as we have argued, only two length scales characterize the solution, there must exist a function, $f(b_+, b_-)$,  with dimensions of length squared, which depends only on $b_\text{ext}^2$ and $\Delta$.
It is reasonable to assume that
 \begin{equation}
 \label{eq:f}
 f(b_+,b_-)=\sum_{m \in S} c_m b_\text{ext}^m\Delta^{2-m}~,
 \end{equation}
where $S$ is some countable set.
To determine the form of $f$, it is convenient to consider solutions with constant scalars.
It is not hard to see that one can have a constant scalar solution if the right hand side of~\eqref{att3} is zero.
In other words~\eqref{A1} must be satisfied and $\phi^i = \phi_\text{ext}^i$.
With constant scalars, our action,~\eqref{actiongen}, essentially simplifies to Einstein--Maxwell gravity and the black hole background reduces to the Reissner--Nordstr\"om solution.
In particular, integrating~\eqref{att2} is easy, and after an simple linear shift of the radial coordinate, one can take 
\begin{equation}
b(r)=r \label{eq:b}~,
\end{equation} so that:
\begin{equation}
\label{eq:rn}
ds^2= - \frac{(r-r_+)(r-r_-)}{r^2} dt^2 + \frac{r^2}{(r-r_+)(r-r_-)}dr^2 + r^2 d\Omega_2^2 ~.
\end{equation} 
Consequently, \eqref{eq:f} then becomes
 \begin{equation}
 \label{eq:f2}
 f(r_+,r_-)=\sum_{m \in S} c_m b_\text{ext}^m\Delta^{2-m}~.
 \end{equation}
Evaluating the Hamiltonian constraint,~\eqref{constraint}, and using~\eqref{Sext}, gives  
\be
r_+ r_- = V_\text{eff}(\phi_\text{ext})=b_\text{ext}^2~.
\label{eq:rprm}
\ee
The important feature of~\eqref{eq:rprm} is that it is independent of the non-extremality parameter, $\Delta$, and we can read off the only non-zero coefficient of~\eqref{eq:f2}: $c_2=1$. Furthermore from~\eqref{eq:rprm} we conclude that $f(b_+,b_-)=b_-b_+$, so that in general, from \eqref{eq:f}, we have:
\be b_+ b_- = b_\text{ext}^2~,\ee
from which the area law easily follows.
As discussed in~\cite{1410.3478}, one can also prove the area law directly using the first law of black hole thermodynamics and the equations of motion or more formally from  the extremization of the field configurations in the region between the two horizons.

Let us define Region 2 as the spacetime between the two horizons, where the radial coordinate $r\in [r_-,r_+]$.
(Region 1 and Region 3 are, respectively, $r<r_-$ and $r>r_+$.)
We summarize the the hot and cold attractor conditions in Table~\ref{table:comp}.
\begin{table}[htb]
	\centering
\begin{tabular}{@{}ll@{}}
	\toprule
			\bf Cold attractors & \bf Hot attractors\\ \midrule
		Decoupled horizon & Coupled horizons \\ \midrule
	Moduli-independent area & Moduli-independent product of areas \\ \midrule
	$\left<\frac{V_\text{eff}(\phi)}{b^2}\,-\,1\right>_{\mathrm{AdS}_2 \times \mathrm{S}^2}=0$&$\left<\frac{V_\text{eff}(\phi)}{b^2}\,-\,1\right>_\text{Region 2}=0$\\[1ex] \midrule
	$\left <\frac{\partial_{\phi^i} V_\text{eff} (\phi)}{b^2}\right>_{\mathrm{AdS}_2\times \mathrm{S}^2} =0$&$\left <\frac{\partial_{\phi^i} V_\text{eff} (\phi)}{b^2}\right>_\text{Region 2} =0$\\ \bottomrule
\end{tabular}
\caption{Comparison of hot and cold attractors}
\label{table:comp}
\end{table}

We conclude from Table~\ref{table:comp} that the attractor geometry equivalent for hot black holes is simply the region between the horizons and all conditions on the scalar flows can expressed as averages over the attractor region.
This establishes Region 2, the inter-horizon region, as the correct geometry to look at when exploring the representations of the microscopic states for hot black holes.\footnote{
The averages for the hot solutions here denotes the inter-horizon radial coordinate average, $\langle F(r)\rangle=\frac{1}{2\Delta} \int_{r_-}^{r_+}dr F(r)$, while those for the cold solutions denote a regulated average over the decoupled AdS$_2$ near-horizon geometry.
The latter is  trivial  as all quantities under consideration are constant all over the attractive geometry.}

\section{Microstate organization}\label{sec:three}
The degrees of freedom living on the horizon of the extremal  black hole give rise to the holographic extremal black hole entropy and are organized in a representation of the asymptotic symmetry group of the attractor AdS$_2$ geometry. We would now like to explore the degrees of freedom of the non-extremal black hole. 

In the last section, we saw that the moduli of hot attractors satisfy an inter-horizon average of the attractor equations for extremal black holes.
For extremal black holes we know that the decoupled near-horizon AdS geometry plays a crucial role in the attractor mechanism. For the hot attractors, we do not have a decoupled\footnote{In other words, there does not seem to be some scaling limit in which one can zoom in one of the horizons to obtain an independent solution. This is not surprising given that we have to average over the inter-horizon region to recover the hot attractor mechanism.} near-horizon geometry let alone an AdS.

Recent attempts to approach this problem for four dimensional black holes involve identifying symmetries of the underlying string or M-theory that leave the horizon geometry unchanged but convert the asymptotically flat backgrounds to asymptotically conical backgrounds~\cite{cl1,cl2,cg,cgs}. Hence, these symmetries acting on solution space take an asymptotically flat black background to a black hole in an asymptotically conical background with modified warp factors. This new solution lifts to a BTZ black hole in an AdS$_3 \times S^2$ background in five dimensions, and therefore can be viewed as a thermal state in the holographically dual CFT, enabling a leading order microscopic formulation of the entropy via the Cardy formula for a two dimensional CFT~\cite{cardy3}. The conformal symmetry appears here in the solution space as opposed to the geometry in the extremal case. However, one of the caveats of this approach is that the final solution has a conical singularity at spatial infinity in four dimensions, and is no longer a good solution of the supergravity theory. Furthermore, the explicit symmetry operation can be executed only for the simplest of $\mathcal{N}=2$ cubic prepotentials like in the STU model. In the following subsection, we will arrive at a background that is asymptotically AdS$_3 \times S^2$ sans the above limitations and which supports a black hole with the same entropy and temperature as the black hole background of interest.

\subsection{``Equivalent'' solutions}\label{subsec:equiv}
We construct a  black hole background with the same entropy and temperature to the black hole of interest in three steps. 
In mapping one system to another, the underlying microstates may differ, but the thermodynamic quantities that characterize the ensemble remain the same.

In the first step, take a black hole with entropy, $S_{\text{original}}$, temperature, $T_{\text{original}}$, and with the asymptotic values of the scalars at an arbitrary point in  moduli space.
The entropy and the temperature encode the two relevant parameters, as $S_{\text{original}}$ is fully determined by $b_+$ and $T_{\text{original}}$ tells us about $b_-$.
As discussed in the previous section,   there exists a constant scalar Reissner--Nordstr\"om solution carrying the same charges as the original black hole with the moduli fixed at the extremal attractor point \eqref{A1}.\footnote{We assume that the black hole potential has a unique extremum defined in terms of its charges and ignore all lines of marginal stability and moduli dependence resulting thereof.} The  metric of the Reissner--Nordstr\"om solution is given by \eqref{eq:rn} which has parameters $r_\pm$. 
 Using \eqref{eq:rprm}, one sees that the charges carried by the black hole fix  the combination ${r_+r_-}$ since they determine the value of $V_\text{eff}(\phi_\text{ext})$. The charges carried by the Reissner--Nordstr\"om solution do not fix its temperature, which we will take to be $\a^2 T_\text{original}$ for some $\a$.\footnote{The explicit relationship between the temperature and the parameters $r_\pm$ can be read off from \eqref{eq:T} with $b_+=r_+$.}

In the second step, we note that there exists a simple scaling symmetry of the equations of motion \eqref{att1}--\eqref{constraint} at the attractor point:\footnote{We need to be at the attractor point so that the right hand side of \eqref{att3} is zero.}
\begin{eqnarray}
a(r)&\rightarrow& \lambda a(r)~,\label{scale1}\\
b(r)&\rightarrow& \lambda^{-1} b(r)~,\label{scale2}\\
(Q_{ea}, Q_m^a) &\rightarrow&   \lambda(Q_{ea}, Q_m^a)~.\label{scale3}
\end{eqnarray}
Here, \eqref{scale3} refers to a scaling operation on all of the charges of the system.
Under this operation, \eqref{scale2} implies
$S \rightarrow \lambda^{-2} S$, so that if  $\lambda$ is chosen such that $\lambda^{-2}= \frac{S_{\text{original}}}{S_+}$, we get a new Reissner--Nordstr\"om solution with entropy $S_{\text{original}}$. 

Finally, from \eqref{eq:T} and  \eqref{scale1}, we see that the temperature scales like $T \rightarrow \lambda^{2} T$. This means that if we take $\alpha=\lambda^{-1}$, the scaled Reissner--Nordstr\"om solution has the same temperature as our original black hole.

We proceed to construct a black background that has an  asymptotic AdS$_3$ factor.\footnote{
Emergent AdS$_3$ factors are also seen, for example, in~\cite{SheikhJabbaria:2011gc}.}
First, we use another symmetry of the equations of motion that was pointed out in~\cite{1410.3478}. This symmetry is a local version of \eqref{scale1} and \eqref{scale2} with $\lambda(r)=r/\sqrt{r_+r_-}$. The charges are {\em not} scaled. The symmetry  takes an asymptotically flat Reissner--Nordstr\"om solution to a black hole in AdS$_2 \times S^2$ at the same non-extremality parameter, with metric,
\begin{equation}
ds^2= - \frac{(r-r_+)(r-r_-)}{b_\text{ext}^2} dt^2 + \frac{b_\text{ext}^2}{(r-r_+)(r-r_-)} dr^2 + b_\text{ext}^2 d\Omega^2~.
\end{equation}
We note that this is the near-horizon metric of the asymptotically flat Reissner--Nordstr\"om black hole found in~\cite{Maldacena:1998uz} in slightly different coordinates.
It will be convenient to shift the radial coordinate, $r\rightarrow r +\tfrac{1}{2}(r_+ + r_-)$, so that the metric becomes
\begin{equation}
ds^2 = -\frac{r^2-\Delta^2}{b_\text{ext}^2} dt^2 + \frac{b_\text{ext}^2 dr^2}{r^2-\Delta^2}+ b_\text{ext}^2 d\Omega_2^2\label{H1}~.
\end{equation}
We now Kaluza--Klein lift this solution to five dimensions, \`a la~\cite{Castro:2008ms}, where it is a solution of minimal supergravity and is a BTZ black hole in AdS$_3 \times S^2$, with the same temperature and entropy as the original black hole. Under the standard AdS$_3/$CFT$_2$ correspondence, this black hole is a thermal state in the holographically dual CFT and its large charge leading order entropy is given by the logarithm of the asymptotic growth of states in the CFT, encoded in the Cardy formula, and expressed in usual notation as,
\begin{equation}
S_\mathrm{BH} = 2 \pi \left(\sqrt{\frac{c L_0}{6}}+ \sqrt{\frac{c \bar{L}_0}{6}}\right) ~.
\end{equation}
This allows us to write down a microscopic  formula for the four dimensional non-extremal Bekenstein--Hawking entropy. 

In the remainder of this note, we analyze the solution space of the Reissner--Nordstr\"om black hole to investigate the origin of its conformal symmetry.
As we shall see, when we consider fluctuations near the Reissner--Nordstr\"om point in the space of solutions of hot attractors, an AdS geometry reappears. This indicates the potential existence of a special subset of supergravity states  of the black hole at fixed gauge charges  that is organized in the $SL(2,\mathbb{R})$ representations of an AdS$_2$ geometry.   Verifying this would require checking that this AdS background is not just an approximate solution at this order in perturbation theory and is in fact an exact solution.
In what follows, we shall adopt this as a working hypothesis.
 
\subsection{Conformal symmetry in solution space}
 To investigate the   solution space, we first start at the Reissner--Nordstr\"om point in the moduli space that lies at the extremum of the black hole potential, and look at fluctuations about this point. (At the Reissner--Nordstr\"om point, the scalars are constant and assume the attractor values.) It should be noted that the fluctuations we consider are not physical fluctuations of a particular solution ---  we are considering nearby points in solution space labelled by the asymptotic values   of the moduli. 
 
Scalar perturbations on this background were previously considered in Appendix~B of~\cite{1410.3478}.
  At zeroth order, we start with the Reissner--Nordstr\"om background (\ref{eq:rn}),~(\ref{eq:rprm}),
 \begin{eqnarray}
 \label{eq:rn_cmpts}
 a_0^2(r) = \frac{(r-r_+)(r-r_-)}{r^2}~, \quad b_0(r)=r~, \quad \phi_0=\phi_\text{ext}~.
 \end{eqnarray}
Suppose the asymptotic value of a modulus is slightly shifted from the attractor value. We  consider
 a scalar near the attractor value, letting
 \begin{equation}
 \phi = \phi_0 + \epsilon \phi_1 + {\cal O}(\epsilon^2)~,
 \end{equation}
 Neglecting backreaction, the equation for $\phi_1$ becomes
 \begin{equation}
 \label{eq:pert2}
 ((r-r_+)(r-r_-)\phi'_1)'= \sigma^2\phi_1/(2 r^2)~,
 \end{equation}
 where $\sigma^2=\p_\phi^2V_\mathrm{eff}|_{\phi=\phi_0}$ is the coefficient of the first order expansion of the right hand side of (\ref{att3}).
 It is convenient to move the poles of \eqref{eq:pert2} from $\{r_+,r_-,0\}$ to $\{1, -1,-\infty\}$ by substituting
 \begin{equation}
 \label{eq:z}
 z=  \frac{r_+(r-r_-)+r_-(r-r_+)}{r(r_+-r_-)}~,
 \end{equation}
which  gives
 \begin{equation}
 \label{eq:pert3}
 \partial_z(r_+r_-(z^2-1)\partial_z\phi_1)=m^2\phi_1~,
 \end{equation}
 where $m^2$ has the usual interpretation of being the mass of the fluctuation about the extremum of the potential:  $m^2 = \frac12\partial^2_{\phi}V_\text{ext}$.

Using \eqref{eq:rprm} and taking $g^{zz}=b^2_\text{ext}(z^2-1)$, one sees that perturbation equation, \eqref{eq:pert3}, is, in fact  the Klein--Gordon equation for static scalar fluctuations, of mass $m$, in an AdS$_2$ geometry:
\begin{equation}
ds^2 = b^2_\text{ext}(-(z^2-1) )dt^2 + \frac{dz^2}{b^2_\text{ext}(z^2-1)}~.
\label{eq:zmet}
\end{equation}
Even though the spacetime is not AdS$_2$, the fluctuations in solution space see an effective black hole in AdS$_2$ at first non-trivial order in perturbation theory. In retrospect, the appearance of an AdS$_2$ should not be surprising given the $SL(2,\mathbb{R})$ symmetry of hypergeometric equations for the scalar field in this background~\cite{larsen}.

In \ref{subsec:equiv} we reviewed the existence of  exact black hole solutions in AdS$_2 \times S^2$   with metric \eqref{H1}.
 As discussed, these are precisely the solutions that lift up to BTZ in AdS$_3$ in five dimensions and allow for a microscopic counting. 
It is  worth noting that when, $\Delta=0$, as expected, \eqref{H1} becomes AdS$_2\times S^2$, which is the near-horizon geometry of the extremal solution. 
The exact AdS$_2$ solution has the same length scale as in \eqref{eq:zmet}. As a classical rescaling of the time coordinate is simply a coordinate change, one can perform  
the rescaling,  $r\rightarrow z\times \Delta$ and $t\rightarrow t/\Delta$, on \eqref{H1} which gives the metric seen by first order fluctuations \eqref{eq:zmet}. 

Once again, it was noted in~\cite{Maldacena:1998uz} that one can eliminate the temperature dependence of \eqref{H1} but the point for us is that this effectively rescales the mass of the perturbations, $m$ by a factor $\frac{1}{\Delta}$, so that 
 the true mass of the fluctuations, $m_\mathrm{AdS}$ in the background, \eqref{H1} is related to the mass in \eqref{eq:zmet} as 
$m_\mathrm{AdS} = \frac{m}{\Delta}$. 
Hence,  in the  zero temperature limit, corresponding to a vanishing $\Delta$,  an infinitesimal perturbation triggers an infinitely massive fluctuation, and consequently, a mass  gap emerges. From this perspective we understand the extremal attractor mechanism as the zero temperature limiting case of the hot attractors. As the temperature decreases, scalar fluctuations are damped as they become more massive so that moduli approach their attractor value as the inter-horizon region shrinks, and the massless fluctuations are localized in the thin laminar region around the extremal horizon and arrange themselves in representations of an $SL(2,\mathbb{R})$ group while the corresponding AdS$_2$ appears as an exact solution of the equations of the motion in the near-horizon geometry. Hence, the black hole in AdS$_2 \times S^2$ acts as an effective attractor geometry in the solution space and which reduces to the spacetime geometry in the extremal limit.

\section{Conclusions and discussion}\label{sec:four}
Let us briefly summarize what we have accomplished so far.
For an extremal black hole, in order to use the Cardy formula to compute the Bekenstein--Hawking entropy, we must first identify the near-horizon AdS$_2 \times S^2$ geometry, which is decoupled from asymptotic infinity by a deep throat.
The near-horizon attractor geometry contains all of the degrees of freedom that are necessary to account for the entropy.
From a gravitational perspective, we can compute the Legendre transform of the action and evaluate it on-shell in the near-horizon region.
This yields the entropy function for the extremal black hole.
The values of the scalar fields obtained from extremizing this function with respect to them are the attractor quantities, while evaluating the function at its extremum gives the entropy of the black hole.
For a statistical enumeration of the black hole microstates, the near-horizon geometry exhibits a conformal structure and is Kaluza--Klein lifted to form the near-horizon geometry of an extremal BTZ black hole, which is then viewed as a chiral thermal ensemble in the dual CFT.
This justifies the microscopic counting, though, of course, it does not identify the precise states in CFT that are dual to the black hole geometry. 

For a non-extremal black hole in the same theory, we have established a parallel procedure.
In the previous section, using the symmetries of the equations of motion, we have constructed a black hole in AdS$_2 \times S^2$ that has the same entropy and temperature as a non-extremal static four dimensional black hole solution of $\mathcal{N}=2$ supergravity.
Our method applies independently of the black hole's temperature.
In particular, we do not specialize to the near extremal case.
In order to perform this analysis, we have fixed scalars to their attractor values.
This isolates a particular solution in the moduli space.
This geometry is then lifted up along a Kaluza--Klein direction corresponding to a graviphoton gauge field to form BTZ black holes in AdS$_3 \times S^2$ that are solutions to five dimensional ${\cal N}=2$ supergravity.
The lift is only possible when the moduli support a weakly coupled supergravity description.
In this case, we have a thermal ensemble in the dual field theory that enables a microscopic enumeration of the Bekenstein--Hawking entropy for the black hole via the Cardy formula.

We have made use of the fact that because of the \textit{hot attractor mechanism} for non-extremal black holes introduced in~\cite{1410.3478}, there are invariants in moduli space constructed in terms of certain combinations of the black hole potential, its derivative, and the warp factor $b(r)^2$ of the $S^2$ that are then expressed as averages over the region between the inner and the outer horizon.
These are precisely the same combinations that are fixed in the attractor geometry in the extremal case by the equations of motion. 
Additionally, classical fluctuations in the inter-horizon region in a non-extremal black hole cannot propagate to spatial infinity.
There is a decoupled region that exists in the absence of an infinitely deep throat.
This supplies a candidate for the part of spacetime geometry which may encode the leading order entropy of non-extremal black holes.
Again, decoupling is key.

Ultimately, in order to understand the statistical physics that underlies gravitational thermodynamics, we want to identify $e^{S_\mathrm{BH}}$ exact microstates that account for a black hole's entropy.
This is a notoriously difficult problem.
Aside from the two charge D$1$/D$5$ system~\cite{sv2} and the $\frac12$-BPS superstar~\cite{llm,bdjs} where we have the maximum supersymmetry compatible with an incipient black hole solution, we do not know how to answer this question.

An easier problem is simply to enumerate the states.
In order to compute the dimension of the Hilbert space of fluctuations corresponding to the decoupled inter-horizon region, we first construct a Reissner--Nordstr\"om solution which lies at the extremal attractor point in the moduli space with the same near-horizon thermodynamic properties, \textit{viz.}, temperature and entropy, as the original black hole.
We then quantize the $s$-wave scalar field fluctuations in region 2.
These are fluctuations in an extremal attractor-like black hole in an AdS$_2$ background and by a rescaling of the time coordinate, they can be brought to the AdS$_2$ factor of the exact black hole in an AdS$_2 \times S^2$ solution of the equations of motion.
Hence, for the purpose of counting states, we adopt the latter geometry as the effective hot attractor background in the space of solutions as opposed to a spacetime solution as in the extremal case.
One can write an entropy function for this background in parallel to the extremal limit, and on extremizing it obtain the Bekenstein--Hawking entropy.

It is often the case that the entropy of a non-extremal black hole is parametrically larger than the entropy of the extremal black hole obtained from taking the zero temperature limit.
Fortunately, this is not the situation that we find ourselves in when employing the hot attractor mechanism in $\mathcal{N}=2$ supergravity.
The zero temperature limit is a smooth limit in solution space in the following sense.
Because of the area law, we know that $A_\mathrm{ext} = \sqrt{A_+\, A_-}$, and this is an invariant for a family of solutions specified by the non-extremality parameter $\Delta$.
When $A_\mathrm{ext}/\ell_\mathrm{P}^2$ is large --- \textit{i.e.}, when the charges are macroscopic and four dimensional $\mathcal{N}=2$ supergravity supplies a semiclassical approximation of quantum gravity --- to leading order, there is no jump in the number of states as we take the $\Delta\to 0$ limit because in this limit $A_\pm \to A_\mathrm{ext}$.
As $r_\pm$ approaches $r_\mathrm{ext}$, the attractor geometry reduces to the usual deep throat decoupling geometry, and the scalar manifold space reduces to the usual attractive flow space at zero temperature.

Recall that we have shown that the mass associated to fluctuations in the space of solutions about the Reissner--Nordstr\"om point behaves as $m_\mathrm{AdS} = \frac{m}{\Delta}$.
In the zero temperature limit, the spectrum therefore acquires a mass gap.
This means that the entropy of the extremal solution is fully determined by massless excitations in the spectrum.
At non-zero temperatures, massive excitations also contribute to the entropy at leading order.
This situation is similar in spirit to the proposal of~\cite{Castro:2009jf} for the near extremal Kerr black hole in which left moving and right moving modes of CFT$_2$ contribute at the same order to the entropy.
In taking the extremal limit, we isolate the degrees of freedom of a single chiral sector of the theory~\cite{Strominger:1998yg,Spradlin:1999bn,kerrcft,Balasubramanian:2009bg}.

In our analysis, we have exploited the symmetries about a special point in the solution space of the non-extremal black hole where the scalars assume the attractor values.
The AdS$_2$ that appears here with an underlying $SL(2,\mathbb{R})$ symmetry becomes an exact symmetry in spacetime in the extremal limit.
As the hot attractor mechanism recovers the equations of motion of the extremal black hole by an averaging over region 2, there is an intimate relation between the moduli space and spacetime geometry.
The decoupled regime between the horizons acts like AdS$_2$ in essential ways.
We seek to make the $SL(2,\mathbb{R})$ of the decoupled region 2 manifest in spacetime.
Future work develops this correspondence further.
One of the other critical assumptions we have made is that the potential has a unique extremum in the region of moduli space corresponding to a single center solution, and this is justified by the extremum value of the potential being completely independent of moduli.
Under a standard Kaluza--Klein uplift, the attractor geometry can then be embedded as a non-chiral thermal state in AdS$_3$, and its microscopic entropy can be computed. 
 This constitutes a preliminary  step in understanding the microscopic organization of non-extremal black hole states in string theory.

The conjecture of~\cite{Martinec:2014gka,Martinec:2015lka,Martinec:2015pfa} for describing the microstates of non-extremal black holes in the D$1$/D$5$ system involves an effective string whose vibrations delimit the two horizons.
Quantization of the vibrational profiles of this string then yields the black hole entropy.
Indeed, according to~\cite{Martinec:2014gka}, degrees of freedom that reside at the inner horizon are crucial to reproducing known thermodynamics of black holes.
This geometrical picture appears not to have an obvious smooth extremal limit within a single duality frame.
Our approach may circumvent this by focusing on the moduli space which does have a smooth limit  and which defines the solution space of the theory better.

In~\cite{cgs}, a symmetry of M-theory into which $\mathcal{N}=2$ supergravity is embedded is used to generate  solutions with conformal symmetry, which could then be lifted to BTZ black holes.
However, this method is only effective for theories where the duality symmetries are explicitly known, such as those with the simplest cubic prepotentials.
Our techniques may apply to more general four dimensional static solutions in $\mathcal{N}=2$ theory.
Indeed, it is remarkable that each step so far in the effort to understand the entropy of non-supersymmetric extremal black holes has a parallel in the extremal case with the same degree of calculational ability.

In addition to clarifying these issues, there are a number of new challenges we must confront.
Our analysis is restricted to black hole backgrounds with an arbitrary non-extremality parameter but which are at a point in the moduli space corresponding to the attractor values of the scalar fields commensurate with the extremal limit. The scalars are constant and do not flow. An obvious next step is to understand the embedding of black holes at other points in the moduli space in terms of CFT ensembles. 

It is also not generically true that one can always lift a four dimensional solution to a weakly coupled solution in five dimensional supergravity with an AdS$_3$ factor.
Fields such as the dilaton, which is one of the moduli, may be strongly coupled along the flow. 
The Cardy formula arises as the leading term in the high temperature expansion of the partition function~\cite{jn}.
If we start off with a charge configuration that is not compatible with the conditions needed for the Cardy formula to apply, we may be able to perform a duality transformation to map the system to a high temperature regime.
The duality transformation acts on both charges and moduli, and one must also show that at the endpoint of the duality operation, the scalar flows still remain in the domain of weakly coupled supergravity.\footnote{For a detailed discussion on these issues see~\cite{jn} and~\cite{Nampuri:2007gw}.}
This is a non-trivial constraint on the system.

This scheme for exploring non-extremal solutions may additionally supply clues to the organization of states in the Hilbert space of black branes of which little is known so far.
In particular, there exist Nernst brane solutions (see~\cite{Barisch:2011ui,BarischDick:2012gj,Dempster:2015xqa,Errington:2014bta,Goldstein:2014qha}) in gauged supergravity theories that exhibit a zero entropy at zero temperature.
Working in gauged supergravity, we may identify and isolate subsectors of the non-extremal Hilbert space, such as Lorentz-violating Lifschitz sectors that occur as intermediate geometries in attractor flows.
Analyzing each of these subsectors might enable us to fully map out enough states to account for the Bekenstein--Hawking entropy.
This is for future work.

\section*{Acknowledgements} 
We thank Gabriel Cardoso, Samir Mathur, Shiraz Minwalla, Larus Thorlacius, and \'Alvaro V\'eliz-Osorio for comments and discussions.
The work of KG is supported in part by the South African National Research Foundation. 
The work of VJ and SN is supported by the South African Research Chairs Initiative of the Department of Science and Technology and the National Research Foundation.
SN would also like to thank the Riemann Fellowship awarded by the Riemann Center for Geometry and Physics and the FCT fellowship FCT-DFRH-Bolsa SFRH/BPD/101955/2014.

\bibliographystyle{JHEP}
\bibliography{nonlocal}

\end{document}